\RequirePackage{fix-cm}
\documentclass[smallextended]{svjour3}       % onecolumn (second format)
\smartqed  % flush right qed marks, e.g. at end of proof
\usepackage{graphicx}
\usepackage{apj-jour}
\usepackage{aps-bibstyle}
\journalname{Rendiconti Lincei}
\def \rosat{{\em ROSAT}\/}
\def \sax{{\em BeppoSAX}\/}
\def \swift{{\em Swift}\/}
\def \agile{{\em AGILE}\/}

\def \fermi{{\em Fermi}\/}
\def \integral{{\em INTEGRAL}\/}

\def \be {\begin{equation}}
\def \en {\end{equation}}

\begin{document}

\title{The key role of \sax\ in the GRB history \footnote{This paper is the peer-reviewed version of a contribution selected among those presented at the Conference on Gamma-Ray Astrophysics with the AGILE Satellite held at Accademia Nazionale dei Lincei and Agenzia Spaziale Italiana, Rome on December 11-13, 2017}}



\author{Filippo Frontera
}


\institute{F. Frontera \at
              University of Ferrara, Dept. Physics and Earth Sciences, Via Saragat, 1 44122 Ferrara, Italy; and INAF/OAS, Via Gobetti 101, 40129 Bologna, Italy \\
              \email{frontera@fe.infn.it}           %  \\
           }

\date{Received: date / Accepted: date}

\maketitle

\begin{abstract}
Twenty years have already been elapsed from the \sax\ discovery of the first afterglow of a Gamma Ray Burst (GRB) in February 28, 1997. 
Thanks to this discovery, it was possible to unveil the 30 year mystery about GRB origin: GRBs are huge explosions in galaxies at cosmological distances.
Starting from the first GRB detection with Vela satellites, I will review the main results obtained before  \sax, the story of the \sax\ discovery of the GRB afterglow with the consequent cosmological distance determination, the main \sax\ results and those obtained after the \sax\ era until to nowadays with the still open problems and prospects.

\end{abstract}

\section{Introduction}
\label{intro}
Gamma Ray Bursts (GRBs) are among the most intriguing phenomena in the Universe. As it is well known, they are sudden bright flashes 
of celestial gamma--ray radiation, with variable
duration from milliseconds to several hundreds of seconds. In rare cases their time duration extends up to 
thousands of seconds. Most of their emission extends from several keV to tens of MeV, but also GeV emission has been discovered in several of them.
With the current instrumentation their observed occurrence rate is 2--3 per day over the entire sky.
Their arrival time is impredictable as it is impredictable their arrival direction.
When they are on, their brightness overcomes any other celestial gamma-ray source. 

In this paper I will briefly review the main steps of the research on GRBs, focusing mainly on the role played by \sax\ in the understanding of their nature. Extended reviews on the \sax\ role can be found elsewhere \cite{Costa11,Frontera15}.

\section{GRB discovery and main efforts before \sax\ }
GRBs were discovered by chance in 1967 with the spy American Vela satellites, 
devoted  to monitor the compliance of the Soviet Union and of other nuclear-capable states with the 1963 
"Partial Test Ban Treaty".
 The American militaries kept this discovery in their drawers until 1973 
when it was published \cite{Klebesadel73}.
The first questions were: which are their progenitors? At which distance are their sites? Which is the power involved?

The solution of these issues implied  complex observational problems, like the accurate 
localization of the events, to test their positional coincidence with known sources. This was  a tough task in the gamma--ray energy band. A possible solution was the search of GRB 
counterparts at longer wavelengths.
 
\begin{figure*}
\centering\includegraphics [width=0.50\textwidth]{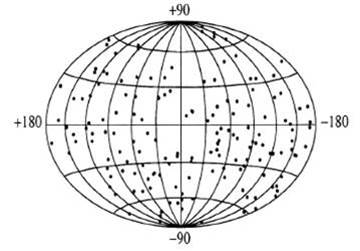}
\centering\includegraphics [width=0.4\textwidth]{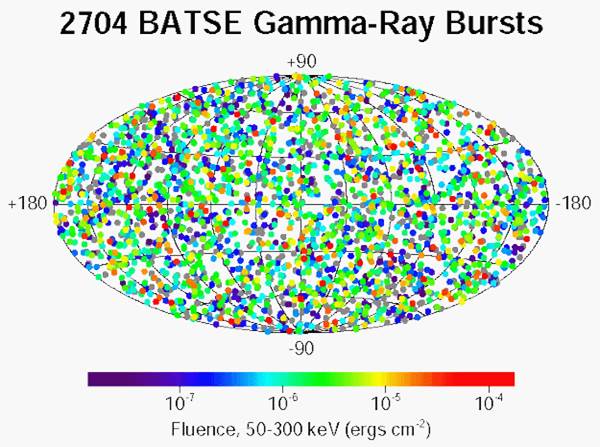} 
\caption{{\em Left}: Distribution of GRBs in the sky in Galactic coordinates as obtained with the Konus experiment  aboard the Venera 11--14 missions. Reprinted from \cite{Mazets88}. {\em Right}: Sky distribution in Galactic coordinates of 2704 GRBs detected with BATSE. Reprinted from \cite{Paciesas99}.}
\label{f:skydistr-Mazets88-Paciesas99}  
\end{figure*}

Many satellite missions in the 70's and 80's (e.g., the Russian satellites Venera 11, 12, 13 and 14, Prognoz 6, Prognoz 9, Konus, Granat, the American Pioneer--Venus Orbiter and the Solar Maximum Mission) included instrumentation devoted to the detection of GRBs (see, e.g., Refs.~\cite{Niel76,Barat81}). The most relevant results about the GRB origin were obtained with the Venera satellites, that found an evidence of an isotropical distribution of the GRB positions in the sky 
\cite{Mazets81b,Mazets88} as shown in the left panel of Fig.~\ref{f:skydistr-Mazets88-Paciesas99}. 

%
% Figure 1
%

The isotropy of the GRB distribution in the sky was later definitely confirmed with the BATSE experiment aboard the CGRO American satellite (see right panel of Fig.~\ref{f:skydistr-Mazets88-Paciesas99}).    

In spite of this result, the sites of the GRBs were still matter of debate, as confirmed by the meeting that took place at the Baird Auditorium of the Smithsonian museum of Natural History in Washington DC (USA) on April 22, 1995 where there were two schools of thought, one led by Bodhan Paczynski and the other led by Don Lamb. The viewpoint of Paczynski was the following \cite{Paczynski95}:"At this time, the cosmological distance 
scale is strongly favored over the Galactic one, but is not proven. A definite proof (or dis-proof) could 
be provided with the results of a search for very weak bursts in the Andromeda galaxy (M31) with an 
instrument ~10 times more sensitive than BATSE. If the bursters are indeed at cosmological distances then 
they are the most luminous sources of electromagnetic radiation known in the Universe. At this time we have no clue as 
to their nature, even though well over a hundred suggestions were published in the scientific journals."
 
Instead the Don Lamb point of view  was very different \cite{Lamb95}: "We do not know the distance scale to gamma-ray bursts. Here I discuss 
several observational results and theoretical calculations which provide evidence about the distance scale. 
First, I describe the recent discovery that many neutron stars have high enough velocities to escape from 
the Milky Way. These high velocity neutron stars form a distant, previously unknown Galactic corona. 
This distant corona is isotropic when viewed from Earth, and consequently, the population of neutron stars 
in it can easily explain the angular and brightness distributions of the BATSE bursts."
  
Thus in 1995 the distance scale of the GRB sites was still a mystery, with the following premonitory conclusion of the debate by Martin Rees \cite{Rees95}: "I'm enough an optimist to believe that it will only be a few years before we know where (and perhaps even what) the gamma-ray bursts are."

\section{The \sax\ revolution}

The great step forward was done with the Italian \sax\ satellite with Dutch participation \cite{Boella97b}: in six months from the beginning of its operational life (September 1996), \sax\ allowed to establish that GRBs are huge explosive events in galaxies at cosmological distances.

Why \sax? Actually the initial main goals of \sax\ did not include GRBs. Only in 1984, two years after the \sax\ approval, during the phase A study, as Principal 
Investigator of the high energy instrument PDS \cite{Frontera97}, I proposed to use the four CsI(Na) scintillator detectors, foreseen as active anti-coincidence shields of the four phoswich detection units,  as 
Gamma Ray Burst Monitor (GRBM) (Internal Report of CNR-TESRE, No. 99, 1984).
The fact that the axis of two of these shields was parallel to that of the two Wide Field Cameras (WFCs)\cite{Jager97}, 
was indeed very suggestive. The expectation was that about 3 GRBs per year (see \sax\ Observers' Handbook, Issue 1.0, 1995) could enter into the common field of view of WFC and GRBM and thus they could be identified as true events by GRBM  and accurately positioned (within 3-5 arcmin) by WFCs. Indeed, transient events detected with WFCs could be originated by other phenomena (e.g., star flaring, X--ray bursts), while their detection in the GRBM energy band was an almost certain signature of GRBs. Obviously it was required  to develop, 
among other things, an in-flight trigger system and a proper electronic chain. The GRBM proposal had also an international echo \cite{Hurley86}.

\begin{figure*}
\centering\includegraphics [width=0.50\textwidth]{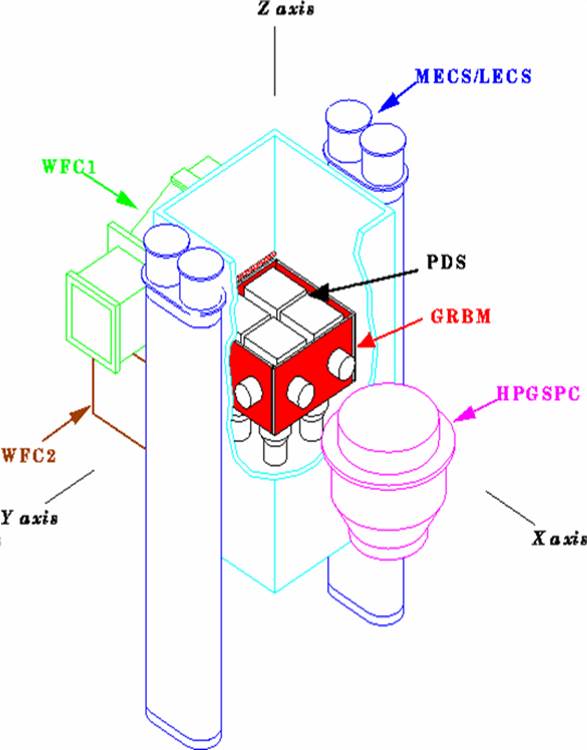}
\caption{The \sax\ payload, with the GRBM shown in red. Note that two opposite GRBM units are oriented as the two WFCs.}
\label{f:sax-payload}  
\end{figure*}

In 1990 a less expensive formulation of the GRBM proposal was approved by the Italian Space Agency (ASI)  
and identified as a further instrument of \sax. The final \sax\ configuration is shown in Fig.~\ref{f:sax-payload} and several descriptions of GRBM were reported 
\cite{Pamini90,Frontera97,Costa98}.

%
% Figure 2
%

The different orientations of the GRBM units implied a
different exposed area to the same GRB. By means of a Monte Carlo
code, we found that this feature could be exploited to obtain a crude GRB direction as later demonstrated with the catalog of GRBs detected with GRBM \cite{Frontera09}. In any case the localization accuracy was sufficient for deriving photon spectra \cite{Pamini90}. In addition, given 
that the GRBM units were located in the center of the satellite, the
implementation of the GRBM  required a very detailed
description of the whole \sax\ satellite based on both simulations and calibrations. 
The simulations were done by developing a very
detailed Monte Carlo code \cite{Rapisarda97}, while the last calibration campaign was performed at ESTEC (Noordwijk, NL)
after the integration of the instrument in the satellite 
\cite{Amati98}.

Before the \sax\ launch, in response to the first international call for the observation of celestial X-ray sources 
with \sax, the PDS/GRBM team submitted a proposal to get WFC data in the case of GRBs identified 
with the GRBM. 
A similar proposal was submitted by the BATSE team for GRBs identified with the 
BATSE experiment. Both proposals were approved.

\sax\ was launched on 30 April 1996 from Cape Canaveral with an
Atlas-Centaure rocket. After a commissioning phase up to June 1996, and a 3 month duration Science Verification Phase (SVP)  
(July--September 1996), the satellite entered in its operational phase in October 1996. 

The first GRB in the field of view of WFC was detected  during SVP, on July 20, 1996, but it 
could be accurately localized only 20 days after the event, and, about one month from the burst, the \sax\ Narrow Field Instruments (NFIs) 
were pointed to the GRB direction: no X-ray counterpart was found \cite{Piro98a}.
 
From this experience, it was clear that a possible residual X--ray radiation, if any, could have been found only in the case it was possible to promptly point the NFIs along 
the direction of well localized events.
To this end, by analyzing the needed operations and
short-cutting all  the interfaces, we set up a
procedure that would minimize the time needed to take a decision about the 
real detection of a GRB by GRBM, its  localization with WFCs, and, in the case a source image was found, request of a \sax\ Target of Opportunity (TOO) to re-point the NFIs toward
the localized GRB. 

\begin{figure}
%\begin{figure}{!h}
\centering\includegraphics [width=0.80\textwidth, angle = 0]{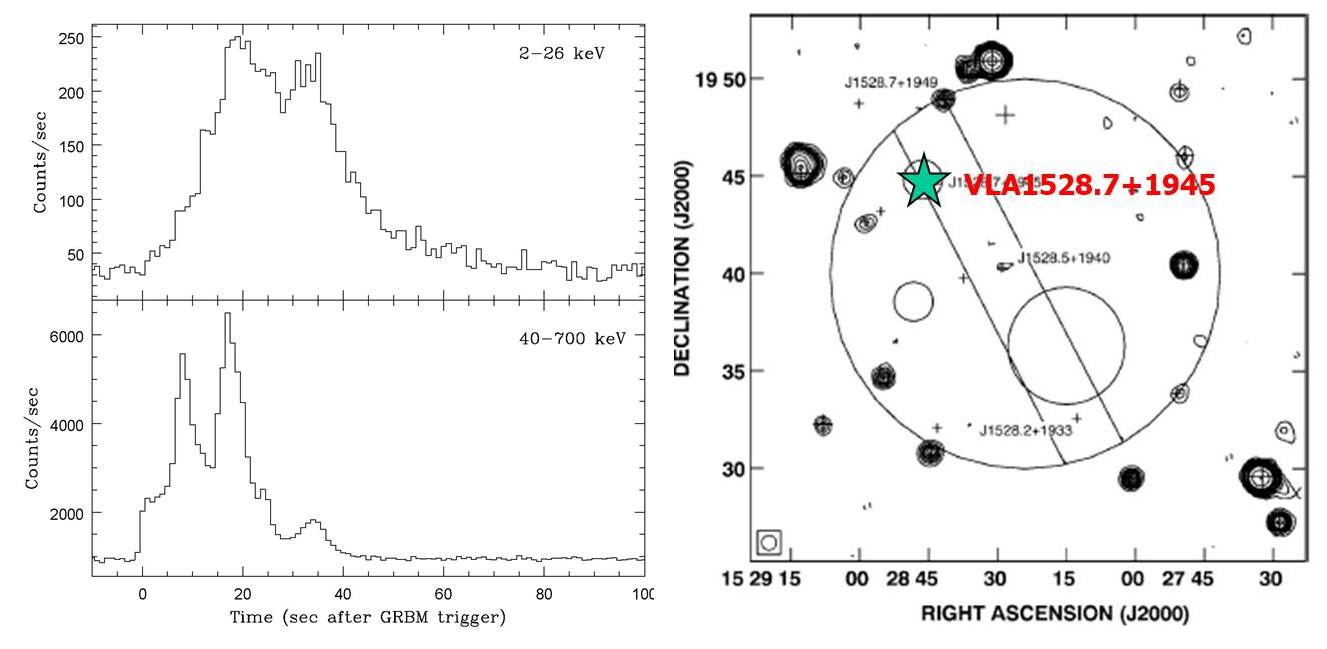}
\bigskip
  \caption{{\em Left}: Light curve of GRB~970111. Reprinted from \cite{Feroci98}. {\em Right}: Earliest and latest WFC error boxes, along with the annulus obtained with the Interplanetary Network. As can be seen, in the refined WFC error box and in the intersection of the WFC error box with the IPN annulus no source was observed. Adapted from the figure of the paper  later published in ApJ\cite{Frail97a}.}
    \label{f:970111-errorbox}
\end{figure}

The first time that the above procedure was entirely applied was on January 11, 1997. A bright burst, GRB~970111, was detected by the GRBM and found in the FOV of one of the two WFCs (see left panel of Fig.~\ref{f:970111-errorbox}). The earliest position of the event was determined with 10 arcmin error radius (see right panel of Fig.~\ref{f:970111-errorbox}). This position was adopted for the X-ray follow-up after 16 hrs.  In this WFC error box a unusual radio source (VLA1528.7+1945), variable on time scales of days, was observed by Dale Frail with the Very Large Array in Socorro (NM, USA).  A similar variability was never observed before in a radio source. The source was thus considered as the possible radio counterpart of the GRB event. A paper was rapidly submitted to Nature. But, after  about 20 days, a revised WFC error box was  produced with centroid about 4 arcmin far from the previous one and with only 3 arcmin error radius (see right panel of Fig.~\ref{f:970111-errorbox}). This new error box excluded the radio source and the paper was withdrawn. The paper was later published in ApJ \cite{Frail97a}. One of the two bright X-ray sources found with the \sax\ NFI was in the same position of the radio source and identified with a non peculiar \rosat\ source \cite{Feroci98} . 
 
%
% Figure 3
%

The experience acquired with GRB~970111 is an important warning for the today multi-messenger astronomy. It shows the need of a very accurate positional coincidence, in addition to the temporal coincidence, in order to have the maximum probability of associating an electromagnetic counterpart to a gravitational wave (GW) or neutrino event, or an optical/IR counterpart to a GRB event. A mission like THESEUS \cite{Amati18}, now accepted by ESA for a phase A study, thanks to its accurate position capability in X--rays and to its prompt follow-up with the IR telescope onboard, is the best solution to optimize the likelihood of the association of a GRB to a GW event and/or an optical/IR counterpart to a GRB. 

\begin{figure}
    \centering\includegraphics[width=0.8\textwidth, angle=0]{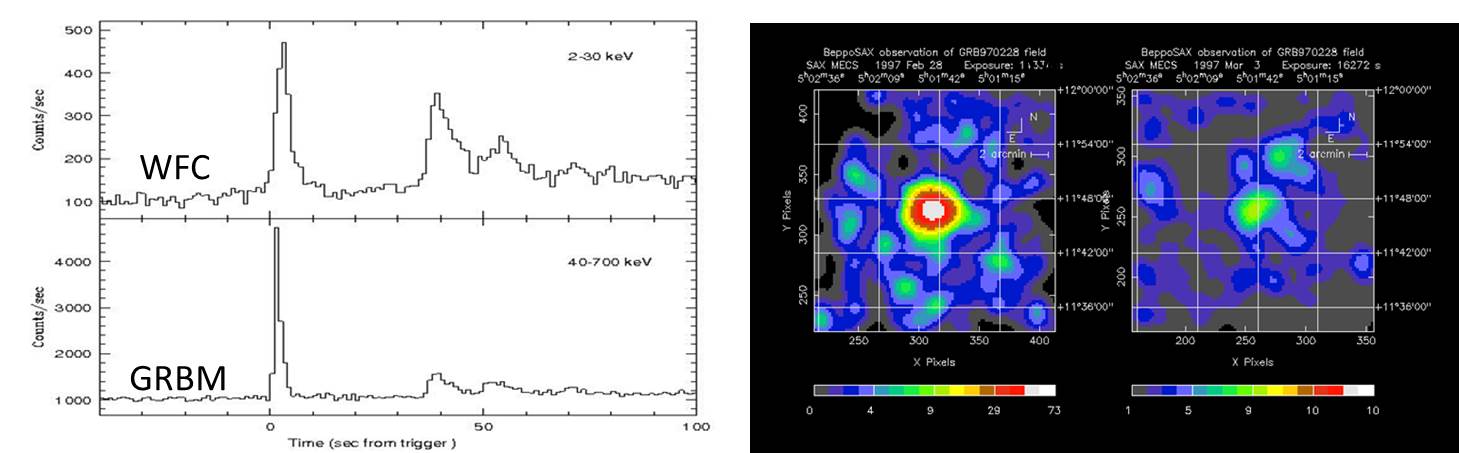}
        \caption{{\em Left}: Light curve of GRB~970228. {\em Right}: X--ray images of GRB~970228 detected with MECS at
        two different epochs. The image on the left includes data from 8 to 16 hours after the burst. The image on the right
 includes 9 hours of data starting 3 days and half after
 the burst. The source was found to be faded by a factor 20.
 Reprinted from~\cite{Costa97}.}
 \label{f:GB0228}
\end{figure}

The solution of the mystery about the GRB sites started with the discovery of the X--ray and optical counterpart 
of the GRB event occurred on 28 February 1997 (GRB~970228).
It consisted of one bright peak, trailed by a train of three more peaks of decreasing intensity (see left panel of Fig.~\ref{f:GB0228}).

The NFI follow-up was performed in 8 hours after the burst time and lasted about 9 hours. In the MECS FOV a previously unknown  source (SAX~J0501.7+1146) was found with 
a flux of $(2.8\pm 0.4)\times 10^{-12}$~erg cm$^{-2}$s$^{-1}$ in 2-10 keV. The field was pointed again three
days after and the source had faded by a factor 20 (see right panel of Fig.~\ref{f:GB0228}).

%
% Figure 4
%

By subdividing the first observation into three subsets, a
light curve was produced. It resulted that the source was decaying according to a power law: $N(t)\propto t^{-1.33}$. 
Quite surprisingly, the flux in the same band detected by the WFC was consistent with the same power law index, 
showing that the fading X-ray source was the delayed emission (afterglow) of the GRB \cite{Costa97}.

\begin{figure}
    \centering\includegraphics[width=0.8\textwidth, angle=0]{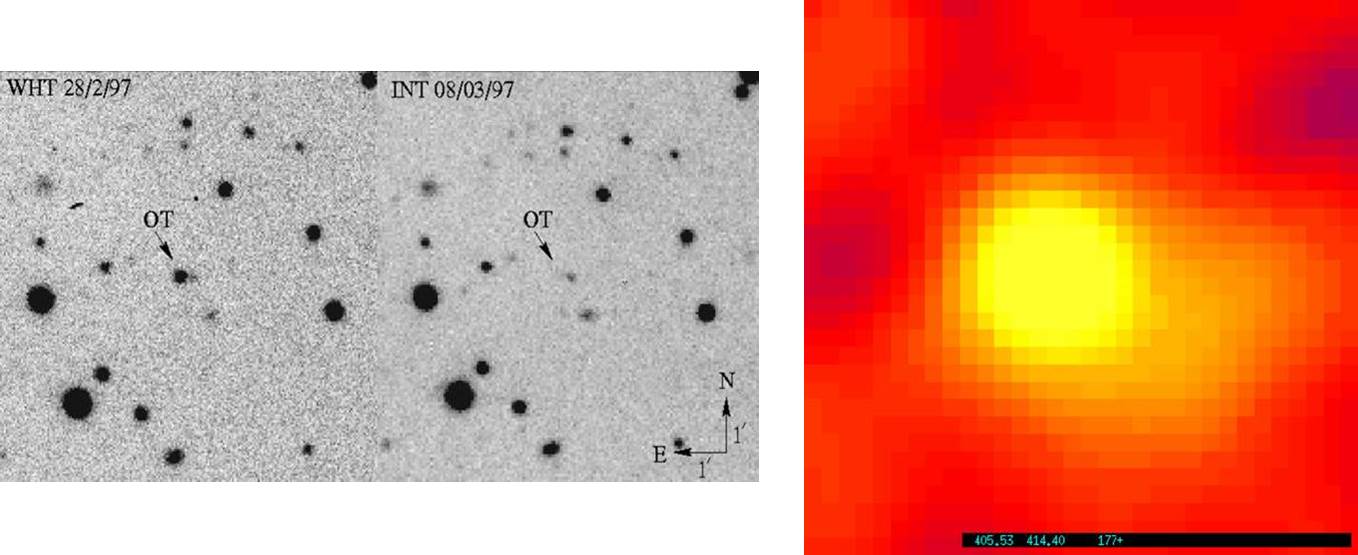}
\caption{{\em Left}: The discovery of the optical counterpart of 
GRB~970228  at two different epochs. The source fading is apparent. See text. Reprinted from~\cite{Vanparadijs97}. {\em Right}: Optical image obtained with HST, performed 39 days after the burst. The point-like source had faded down to V magnitude 26.4 and  was embedded in a faint nebular source with $V\approx 25$ extended $\sim 1$ arcsec, likely, but yet not necessarily, a galaxy. Reprinted from \cite{Sahu97}.}
 \label{f:fig970228-optical}
\end{figure}

Meanwhile, the coordinates of WFC and those improved of NFIs were distributed by the GRBM team 
directly and through IAU circulars (see, e.g., \cite{Costa97a}). Various observers performed
optical observations. The group led by Jan Van Paradijs was the first to
perform two observations of the same field with the same filter.
From the comparison of two images taken with the William Herschel Telescope on February 28 and other two 
taken with the same telescope and with the Isaac Newton Telescope on March 8, it resulted that a previously unknown optical 
point-like source 
was fading (see Fig.~\ref{f:fig970228-optical}) from $V = 21.3$ to $V>$ 23.6 \cite{Vanparadijs97} . 

%
% Figure 5
%

\begin{figure*}
\centering\includegraphics[width=0.8\textwidth, angle=0]{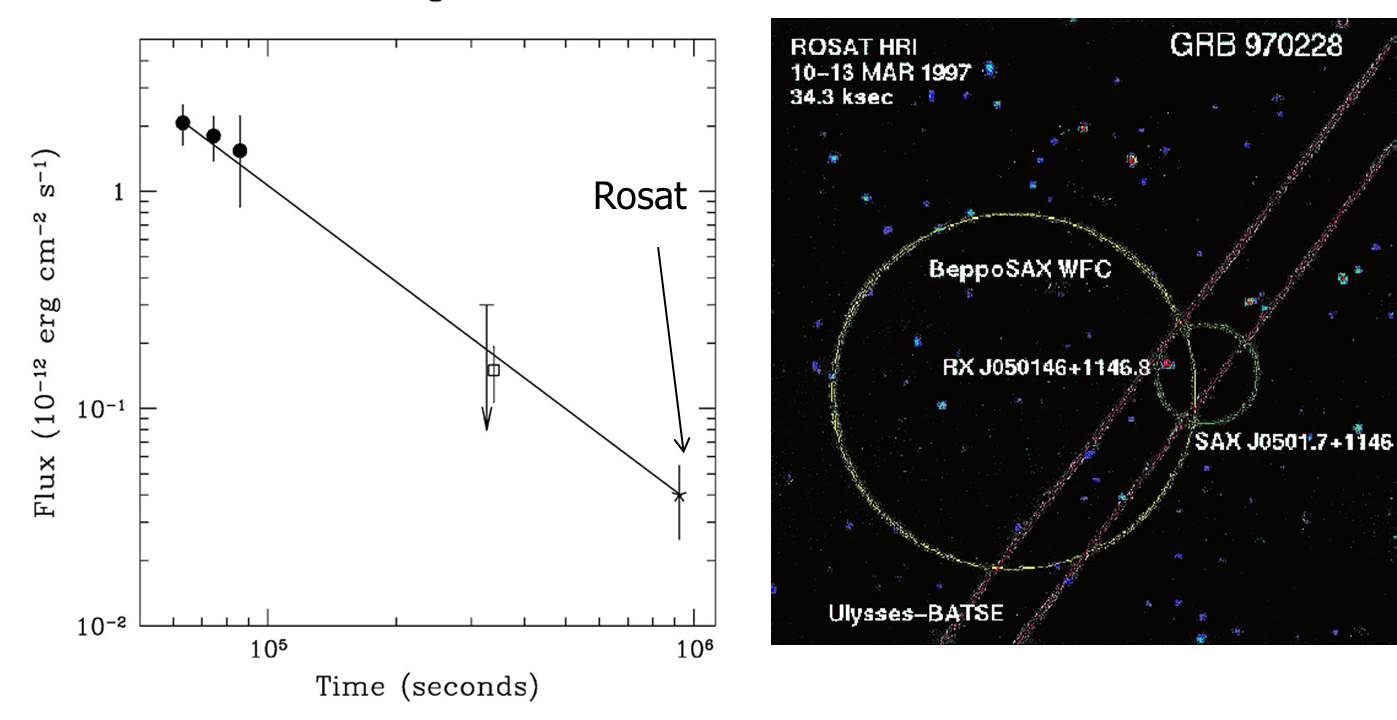}
\caption{{\em Left}: The decline of the 0.1--2.4~keV GRB~970228 flux with time from the burst
onset, uncorrected for Galactic absorption. The filled dots are the LECS
data points, the arrow is the LECS 3$\sigma$ upper limit,
the square gives the flux extrapolated
from the MECS detection, and the cross shows
the \rosat\ HRI data point. The best fit power-law decay is also shown. Reprinted from \cite{Frontera98b}. {\em Right}: The central 8 arcmin image of the HRI field of view of the ROSAT observation on 1997 March 1013. 
The only source above 3$\sigma$ in this image is RXJ050146+1146.8 inside the smaller circle, which is the 1 arcmin 
error circle of the fading source SAX~J0501.7+1146 as found with the \sax\ NFIs. The \rosat\ source is coincident with the optical source (in red) within 2 arcsec. The large circle shows 
the 3 arcmin error circle of GRB~970228 as determined with the \sax\ WFC, while the two
straight lines mark the triangulation circle derived from \sax\ GRBM and {\em Ulysses} timings of GRB~970228. 
Reprinted from \cite{Frontera98b}.}
 \label{f:rosat}
\end{figure*}

To confirm the association of the X-ray afterglow source with the GRB event, we proposed a 
pointing of the event with the High Resolution Imager (HRI) aboard the X--ray satellite \rosat. This instrument, 
in the 0.1--2.4 keV passband, could provide the best position and the smallest error box of the event direction, 
thanks to its high angular resolution (10 arcsec radius) and  sensitivity.  The observation was
performed on March 10 and lasted three days. HRI detected 8 sources in its 20 
arcmin FOV, but only one (RXJ050146+1146.9) was within the error box of the
afterglow source found with \sax\ \cite{Frontera98b} and found  (see right panel of Fig.~\ref{f:rosat}) coincident, within 2 arcsec,  with the discovered optical transient. The intensity of the \rosat\ source was found to be fully consistent (see left panel of Fig.~\ref{f:rosat}) with the power law derived from the \sax\ LECS afterglow spectrum estimated in the 0.1-2.4 keV band \cite{Frontera98b}. 
With these results, the association of the ROSAT and \sax\ sources with the 970228 afterglow became conclusive.

%
% Figure 6
%

Also the spectral analysis of the event and its afterglow was performed \cite{Frontera98a}. While the spectrum of the prompt event was consistent with a Band function and showed, within each peak, the already known hard-to-soft evolution, 
the spectrum of the afterglow
was a power law, $N(E) \propto E^{-2.04}$, constant with
time.  

In conclusion, both temporal and spectral trends of the afterglow advocated in favor of a non thermal 
process and would be, in the following, the basic building blocks for all the GRB theories.

An observation of the Hubble Space Telescope, performed 39 days after the burst, showed that the
point-like source had faded down to V magnitude 26.4 and  was embedded in a faint nebular source (see right panel of Fig.~\ref{f:fig970228-optical}) with $V\approx 25$ extended $\sim 1$ arcsec, likely but yet not necessarily, a host galaxy \cite{Sahu97}.

\begin{figure*}
    \includegraphics[width=1.0\textwidth]{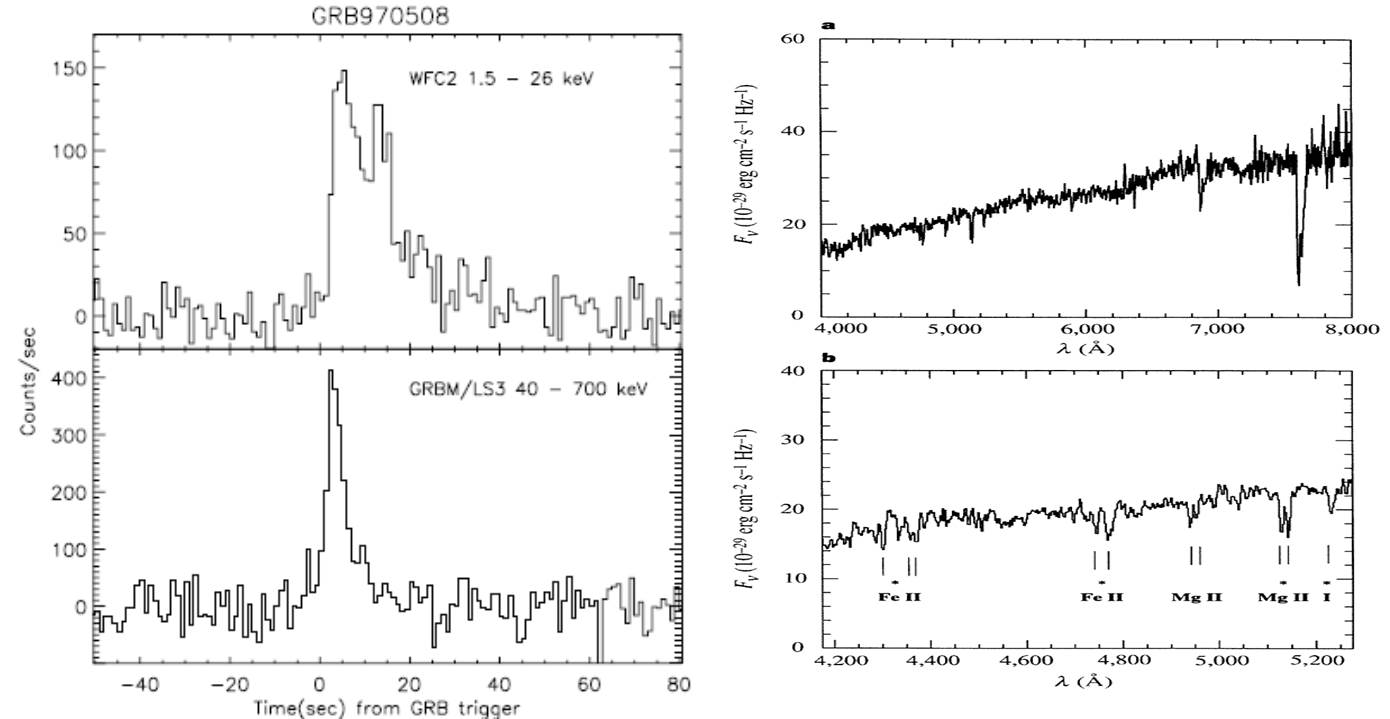}
\caption{{\em Left}:The light curve as observed by the GRBM (40--700 keV) and WFC (1.5--26 keV).  Reprinted from \cite{Piro98b}. {\em Right}: Spectrum of the optical counterpart of GRB~970508 taken with the Keck Low
Resolution Imaging Spectrograph \cite{Metzger97}. Various absorption lines were
found: some at redshift $z= 0.835$, some other at red-shift of $z = 0.767$. The first of these two $z$ values was found, weeks later, in the emission lines from  the galaxy that had hosted the fading object.} 
 \label{f:970508}
\end{figure*}

The further turning point of the \sax\ discovery was on May 8, 1997, when the event GRB~970508 was 
identified by GRBM and localized with WFC (see left panel of Fig.~\ref{f:970508}). The X--ray follow-up with \sax\ NFIs was the fastest up to then: 5.4 hours after the burst. The X--ray afterglow was detected. The NFI position 
was soon distributed and an optical counterpart was detected. The new object, contrary to 
the GRB~970228 optical counterpart, had a flux that increased for around two days, 
arriving to $R = 20.14$ and then started to fade with the usual power law. On May 11, 
when the optical afterglow was still relatively bright, 
the CalTech/NRAO group observed it with the Keck Low
Resolution Imaging Spectrograph. Various absorption lines were
identified: some at red-shift $z= 0.835$, some other at red-shift of $z = 0.767$  \cite{Metzger97}. 
The first of these two $z$ values was found, weeks later when the point--like object had 
almost completely faded, in the emission lines from  the galaxy that had hosted the fading object. 
The mystery of the GRB sites was solved.  Remote galaxies harbour GRBs.

%
% Figure 7
%

The immediate consequence was to fix the energy scale.  From the derived luminosity distance we derived the energetics 
of GRB~970508: $E_{iso} = (0.61\pm 0.13) \times 10^{52}$ ergs, assuming isotropic emission.

GRB~970508 was also relevant for the discovery of the first
radio afterglow with the VLA radio telescope \cite{Frail97}. The radio emission showed the phenomenon known as
\textit{scintillation} that derives from the effects of
interstellar clouds on sources of very small angular size. In
GRB~970508 the scintillation disappeared after around two months.
From the angular size and from the distance, Frail et al. \cite{Frail97} derived the expansion 
velocity of the fireball that came out to be around $2c$, an apparent superluminal effect 
typical of sources expanding at relativistic velocity.

\section{International resonance of the \sax\ discoveries}

The resonance of the \sax\ discoveries in the scientific community was enormous. In the first two years (1997-1998) the number of papers citing 
\sax\ was equivalent to that citing HST (about 200/yr). For two years the GRB discoveries where classified by the Science journal among the top ten over the world and over all the science fields.

ESA modified the data flow of the INTEGRAL satellite in order to allow a prompt localization of GRBs with the on-ground analysis of the gamma--ray imager IBIS data.

NASA issued an Announcement of Opportunity for a new medium-size scientific satellite: many missions dedicated to GRBs were submitted and one (\swift, now Neil Gehrels \swift) selected.
The \swift\ satellite has the same \sax\ configuration, but the GRB localization and X--ray telescope re-pointing are automatically performed in a very short time ($\sim 100$~s) \cite{Gehrels04}. 

The largest radio and optical telescopes devoted observing time to follow-up GRBs localized with \sax. Some of them modified their procedures or their equipment to make these observations faster.

Several optical or NIR telescopes were built with robotic pointing of the coordinates distributed by \sax\ through the already existing GCN network set up by NASA, which got an impressive boost by the \sax\ findings.

In addition, the \fermi\ satellite was designed taking into account the \sax\ payload configuration: a GBM instrument to identify GRBs, LAT telescope to localize them \cite{Atwood09,Meegan09}. A similar design was adopted for \agile, with a gamma-ray imager sensitive in the range 30~MeV-- 50~GeV, and a hard X-ray imager (Super\agile) sensitive in the range 18-60 keV with a Field of View (FOV) of about 1 sr \cite{Tavani08}.

\section{Immediate consequences of the \sax\ discoveries on the GRB theoretical models}
\label{s:theory}

\begin{figure*}
	\centering\includegraphics[width=0.8\textwidth]{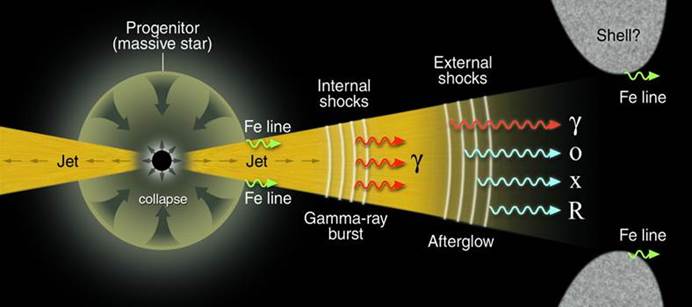}
		\caption[]{The schematics of a relativistically expanding fireball. Due to a super--Eddington luminosity, internal energy is converted in kinetic energy, which, through shocks (internal, external),  is transformed in electromagnetic energy. Figure adapted from \cite{Meszaros14}. }
	\label{f:fireball}
\end{figure*}

The cosmological distance scale of GRBs  swept away all the Galactic models. 
The observed properties, like an energy release in  gamma-rays up to $\sim 10^{54}$~erg (assuming isotropy) in a short time (tens of second), the non-thermal spectra, the short time variability (down to ms time scale), the photon energies $>$ 1 MeV, were generally interpreted as a result of the formation of a fireball in relativistic expansion (see Fig.~\ref{f:fireball}). This model, already developed before the \sax\ discovery of the X--ray afterglow (e.g., \cite{Guilbert83,Goodman86,Paczynski86}), had an immediate success for its capability
to explain the spectral and temporal GRB properties (e.g., \cite{Wijers97,Sari98}), through the conversion of the fireball kinetic energy into electromagnetic radiation. This conversion was assumed to occur 
through shocks between contiguous shells within the fireball for the prompt emission, or with the external medium 
for the afterglow emission (see, e.g.,\cite{Meszaros94,Paczynski94}). 

In spite that the fireball model was considered the GRB standard model, some drawbacks were noticed. Among them, the small conversion efficiency of the internal shocks (e.g., \cite{Daigne98}), while external shocks should be more efficient. This was found inconsistent with the observation results: more energy released during the prompt emission, at least on the basis of the afterglow measurements available up to 10 keV.

%
% Figure 8
%

Concerning progenitors,  several models were proposed. For short GRBs, the merging of a white dwarf--neutron star system, or that of a binary neutron star system or that of a neutron star--black hole system were considered the most likely mechanisms  \cite{Narayan92}. For long GRBs, instead, failed supernovae \cite{Woosley93} or the collapse of a rapidly rotating star to a Kerr black hole (collapsar) \cite{Paczynski98} with the formation of hypernovae, were the most favorite models. But also other models were proposed, like the 
supranova model \cite{Vietri98}, the transition, by accretion,  of a neutron star to a deconfined quark star \cite{Berezhiani03}, the ElectroMagnetic Black Hole (EMBH) model \cite{Ruffini01}.

\section{Some major further \sax\ discoveries on GRBs}

Other relevant results were obtained with \sax. They include:

a) the discovery of a transient absorption edge at 3.8 keV in the prompt emission of the \sax\ GRB~990705 \cite{Amati00}. The feature was found to be  consistent with a redshifted K-edge of an Iron environment. The derived red-shift ($z = 0.86$) was later found to be consistent with that measured from the GRB host galaxy \cite{Lefloch02}.

b) the discovery of a decreasing column density in the prompt emission from the \sax\ GRB~000528 \cite{Frontera04a};
 
c) Discovery of the GRB/Supernova connection. The location of the \sax\ GRB~980425 was found to be consistent with that of the type Ic supernova  SN1998bw explosion \cite{Galama98b}. Beside the positional coincidence, the SN explosion was simultaneous, within one day, with GRB\,980425, and, thence, the latter was the likely starting event. The SN was unusually bright (hypernova) and characterized by a high expansion velocity \cite{Patat01}.  The uncertainty about the chance coincidence 
of SN1998bw with GRB\,980425 was definitively removed in 2003, 
when the type Ic SN2003dh was found to be associated with GRB\,030329 \cite{Stanek03}. Indeed, while the 
early spectra of the optical emission  from SN2003dh consisted of a power-law continuum, after a week, these spectra, corrected for the afterglow emission, became remarkably similar to that of SN1998bw. 
Nowadays it has been definitely confirmed that several long GRBs orginate in supernova explosions (see Table~\ref{t:SNe-GRBs}). In general these 
SNe are type Ic with high expansion velocities and much larger energy release  than  in  normal SNe. 
%From radio observations, it was also found that less than 3\% of %local SNe Ibc show evidence for association with a GRB or X-ray %flash (XRF, see below) \cite{Soderberg07}.
However there are GRBs not associated with SNe (see, e.g., \cite{DellaValle06a}), demonstrating that there are GRBs originating in very faint supernovae or they are due to different phenomena (see Table~\ref{t:SNe-GRBs}). For a recent review on GRB-SN connection, see \cite{Hjorth13}. 

\begin{figure}
	\centering\includegraphics[width=0.8\textwidth]{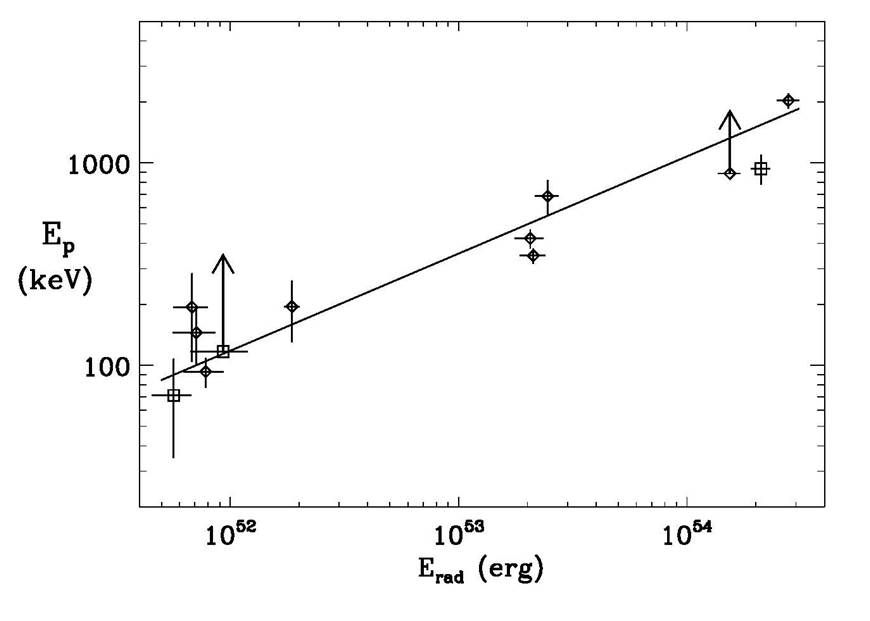}
		\caption[]{The $E_{p,i}$--$E_{iso}$ relation discovered with \sax. Reprinted from \cite{Amati02}.}
	\label{f:ep-eiso}
\end{figure}

%
% Table 1
%
\begin{table}
\caption{List of the SNe associated with GRBs. Most of them have a known redshift with values $<1$. As can be seen, there are also cases in which, in spite of a low redshift, there is non evidence af an associated SN.}
	\centering\includegraphics[width=1.0\textwidth]{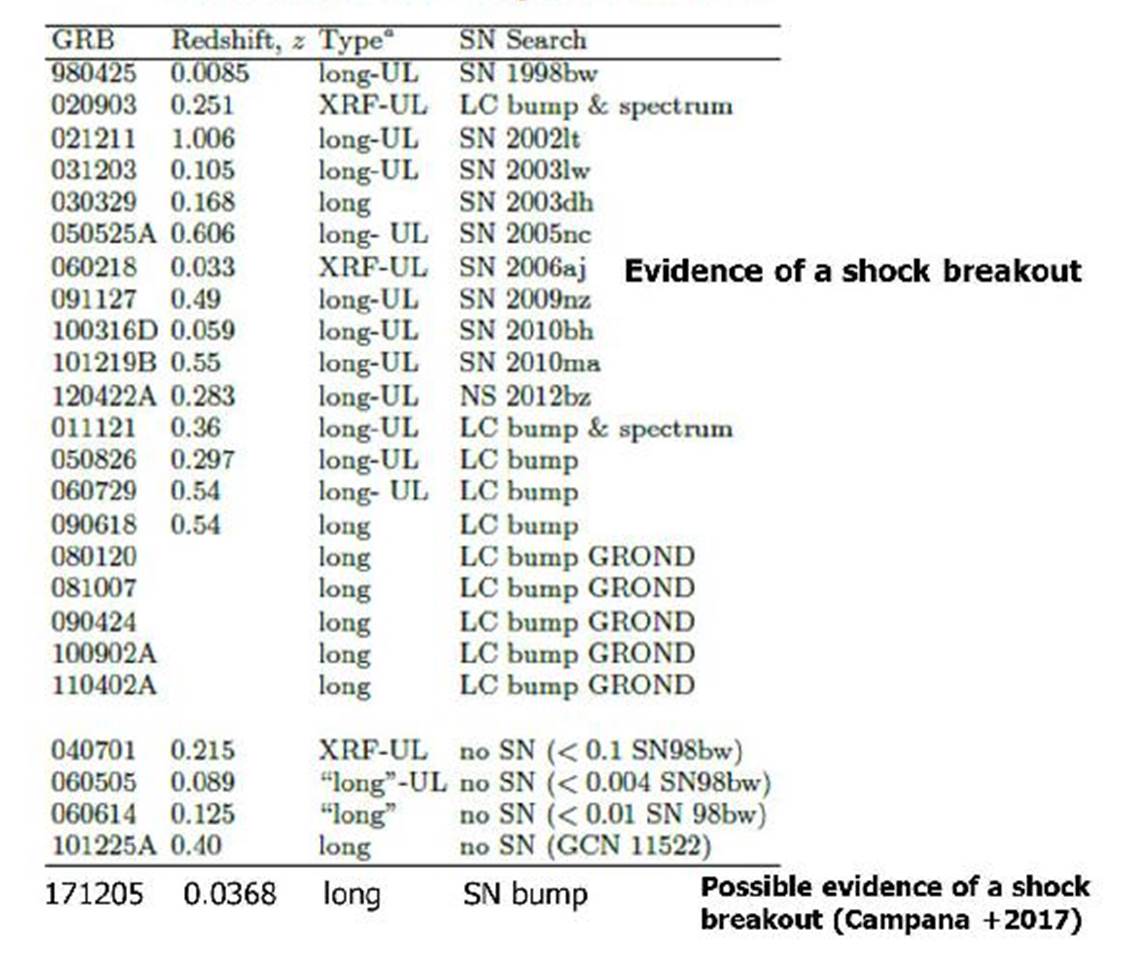}		
	\label{t:SNe-GRBs}
\end{table}

e) Discovery of the "Amati relation".
This relation concerns the  correlation between the 
redshift-corrected photon energy $E_{peak}$, at which 
the $\nu F_\nu$ spectrum peaks, and the total released energy during the burst $E_{iso}$ 
in the hypothesis of isotropic emission (see Fig.~\ref{f:ep-eiso}). It was found with a set of \sax\ GRBs, whose redshift was determined with optical spectrometers \cite{Amati02}. After we published this result, other relationships were reported: the "Yonetoku relation" between $E_{peak}$ and the bolometric peak luminosity $L_{p,iso}$ \cite{Yonetoku04}, the "Ghirlanda relation" between $E_{peak}$ and the released energy $E_\gamma$ corrected for the beaming factor ($E_\gamma = (1- cos\theta)E_{iso}$), given that a jet-like structure for the GRB emission was assumed \cite{Ghirlanda04a}. A discussion of the weaknesses of the Ghirlanda relation can be found in \cite{Frontera12a}.  Other relations were later reported between prompt and afterglow emission or concerning only the afterglow. For a review see \cite{Dainotti17}. 
 In spite of the more recent relations, the Amati relation remains the most robust and it is now confirmed by all long GRBs for which it has been possible to derive, 
along with $z$, their bolometric fluence and peak energy $E_p$. The only exception is the nearby GRB~980425. Some authors suspected that this relation 
could be influenced by selection effects (see, e.g.,\cite{Butler09}). However, the time 
resolved spectra, obtained by slicing the GRB time profile in several time intervals and 
deriving the spectra in each of them, show a correlation between the time resolved  $E_p$ and the
corresponding flux (see, e.g., \cite{Ghirlanda10,Frontera12a,Frontera12b}). Thus it is now generally accepted. A possible interpretation of the correlation has also been recently proposed by us \cite{Frontera16}. Thanks to this relation, GRBs appear to be a promising tool to describe the expansion rate history of the universe and an indepentent  estimate of the cosmological parameters (see, e.g., \cite{Amati13}).

e) Discovery of X--ray flashes. These events were detected with the WFCs aboard \sax\ in the 2--25 keV energy band as bright X--ray sources lasting of the order of minutes, but remaining undetected in the \sax\ GRBM \cite{Heise01}. Their temporal and spectral properties were found very similar to those of the X--ray counterparts of GRBs.

%
% Figure 9
%

\section{The post-\sax\ era}

In spite of the huge advances obtained thanks to the \sax\ discoveries, many questions about
GRBs were left unanswered, like the early afterglow properties, the late breaks in the X--ray light curves, the afterglow of short bursts, the origin of dark GRBs, the GRB environment,
the origin of X--ray flashes.

\begin{figure}
	\centering\includegraphics[width=0.8\textwidth]{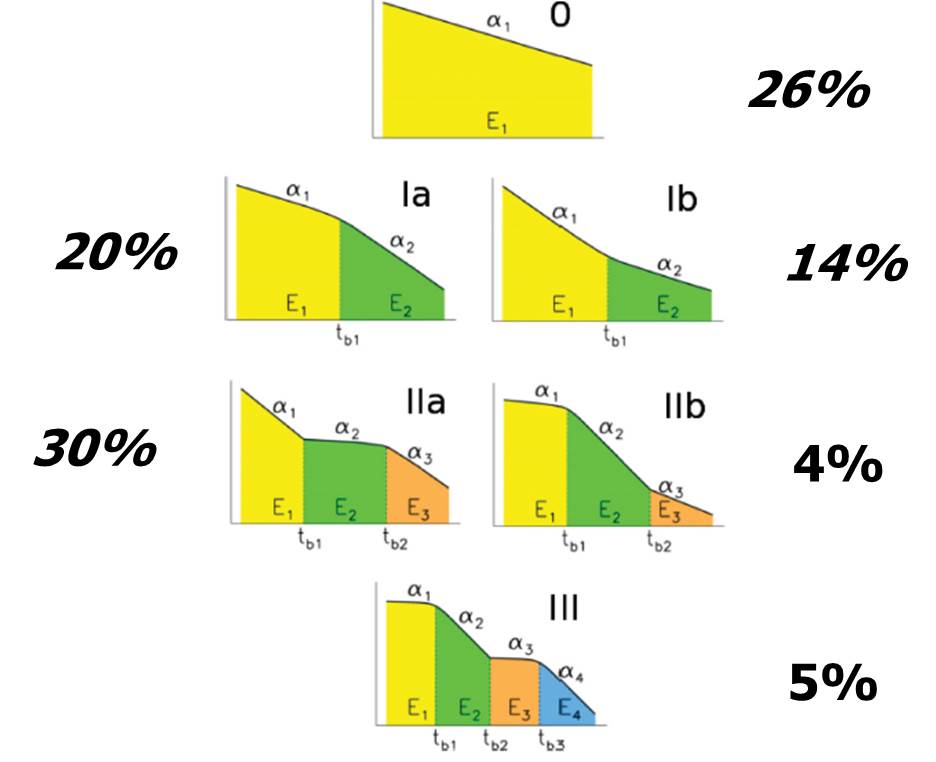}
		\caption[]{The different light curve models of the early X--ray afterglow as observed with \swift. Also the occurrence frequency is shown. Initially it seemed that the light curve canonical model was the IIa, which only occurs 30\% of the times. Figure adapted from that of \cite{Margutti13}.}
	\label{f:early-afterglow}
\end{figure}

\begin{figure}
	\centering\includegraphics[width=0.8\textwidth]{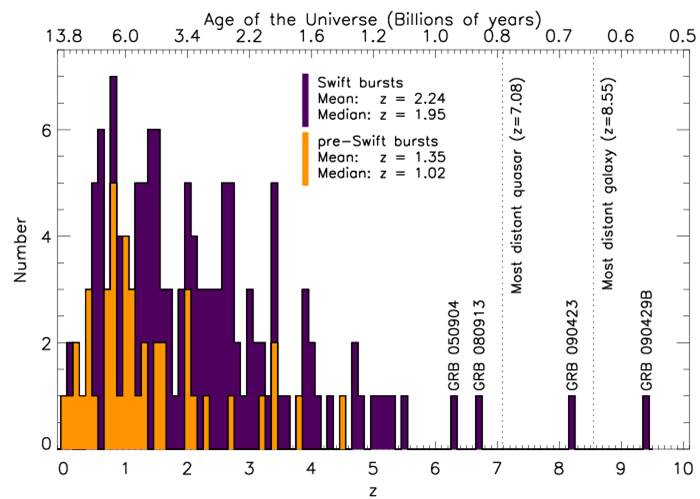}
		\caption[]{Redshift distribution of GRBs. Most of them have been discovered and localized with \swift. Reprinted from the review by \cite{Gomboc12}}
	\label{f:z-distribution}
\end{figure}

One of the missions that is giving a very high contribution to the GRB astrophysics is \swift, launched on November 20, 2004.
Several results have been obtained and reviews of the most important ones  have been reported (see, e.g., 
\cite{Gehrels09,Kumar15,Gehrels17}).
I wish to mention here some of these results, like the different decay modes \cite{Margutti13}  of the early X--ray afterglow light curves (see Fig.~\ref{f:early-afterglow}), and the great contribution to the determination of the redshift distribution of GRBs (see Fig.~\ref{f:z-distribution}).

%
% Figure 10
%
%

%
% Figure 11
%

\begin{figure*}
	\centering\includegraphics[width=1.0\textwidth]{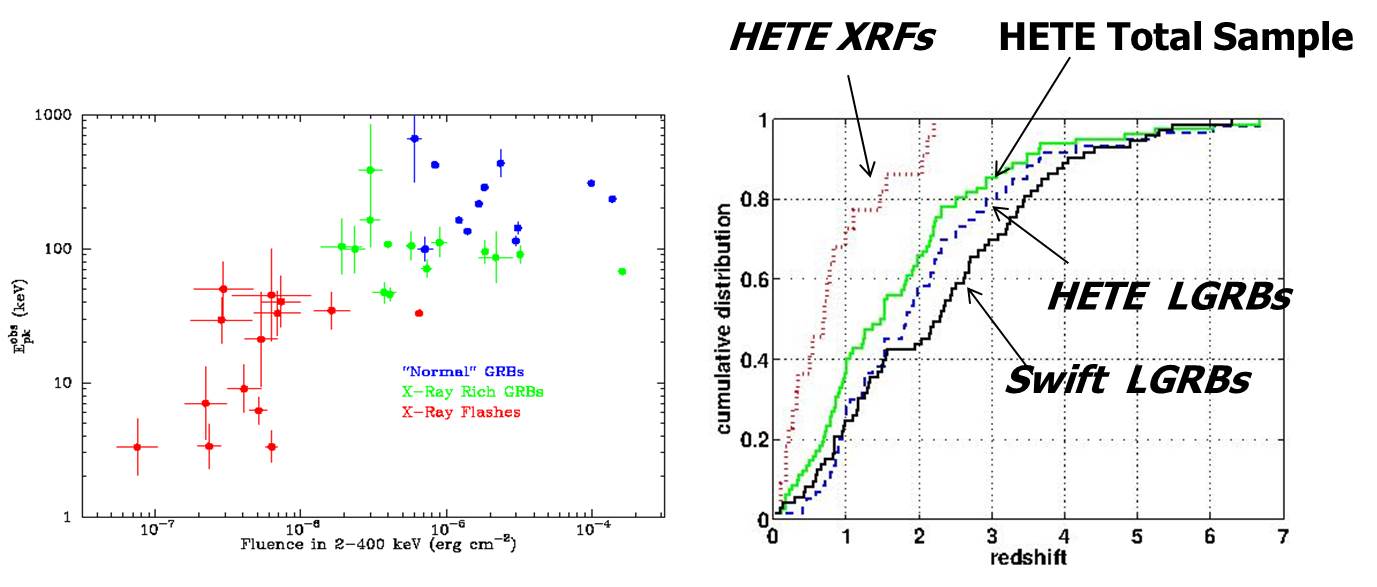}
		\caption[]{{\em Left}: Distribution of the peak energy $E_p$ with fluence separated for different classes of events (X--ray Flashes, X--ray Rich, GRBs). Adapted from \cite{Sakamoto05}.  {\em Right}: $z$ distribution of GRBs and X--ray Flashes. Figure adapted from that of \cite{Pelangeon08}.}
\label{f:xrayflashes}
\end{figure*}

Also the HETE~2 mission (e.g., \cite{Lamb04a}), launched in October 2000, had an important role in the post-\sax\ era, in particular for the understanding of the X--ray flashes (see, e.g., \cite{Pelangeon08}). Thanks to HETE~2 it was possible to establish that X--ray flashes show properties similar to those of GRBs, apart their lower peak energy (see left panel of Fig.~\ref{f:xrayflashes}) and a lower $z$ distribution (see right panel of Fig.~\ref{f:xrayflashes}).

%
% Figure 12
%

\begin{figure*}
	\includegraphics[width=1.0\textwidth]{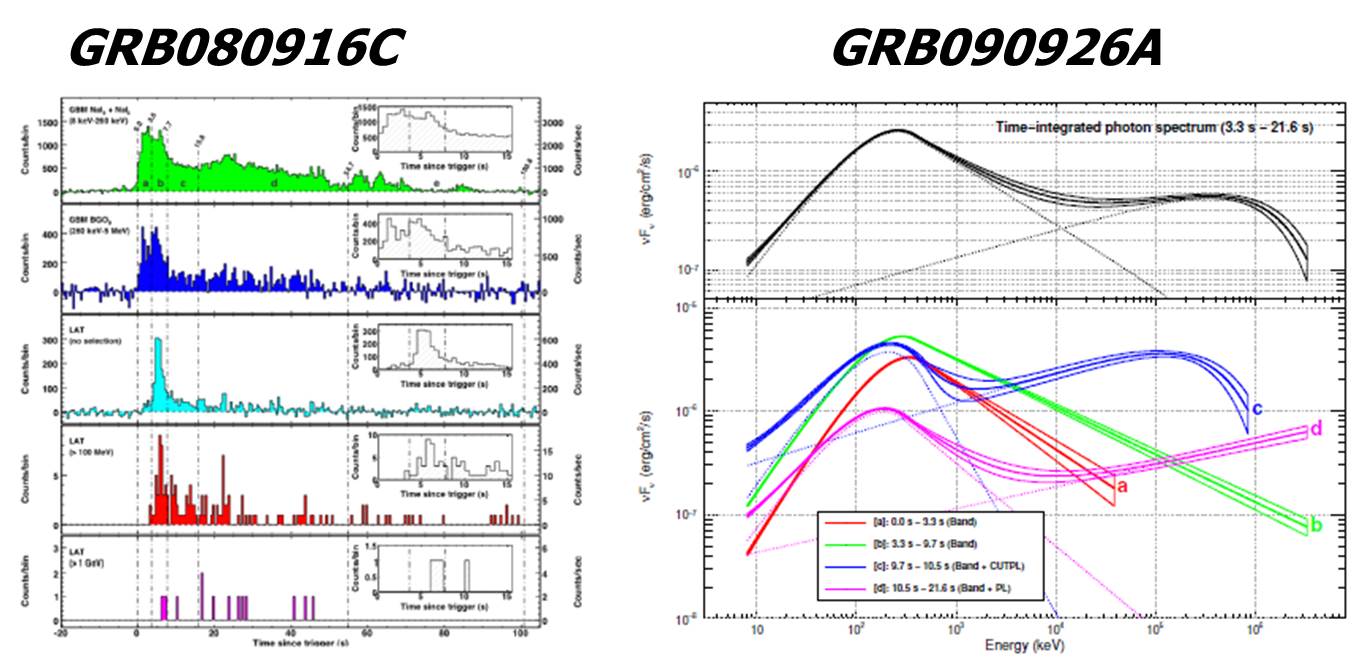}
		\caption[]{{\em Left}: Light curve of the \fermi\ GRB~080916C at different energies. As can be seen, the onset of the high energy gamma--rays delay with respect to those at low energies. Reprinted from \cite{Abdo09a}. {\em Right}. Time resolved broad--band spectra of the \fermi\ GRB~090926A. A hardening of the high energy spectrum with time is apparent with a gamma--ray spectral component visible during the latest time interval. Reprinted from \cite{Ackermann11}.}
\label{f:high-energies}
\end{figure*}

Also the \fermi\ mission launched on June 11, 2008  and the \agile\ mission launched on April 23, 2007 are providing a great contribution to the understanding of the GRB phenomenon at  high energies (see, e.g., \cite{Zhang11}).  One of the most intriguing results is the delay of the onset of the high gamma--ray energy light curves with respect to that at low energies (see left panel of Fig.~\ref{f:high-energies}), and the hardening of the spectrum with time from the GRB onset with the appearance of a high-energy spectral component in the tail of the gamma--ray light curve (see right panel of Fig.~\ref{f:high-energies}).
 
%
% Figure 13
%

Concerning progenitors, there is a general consensus that long GRBs  are the result of core collapse of very massive stars. This conclusion comes from the following facts: a) the well established GRB--SN connection; b) long GRBs are located in the brightest regions of the galaxies with high star formation rate (SFR) and where the most massive stars occur. 

Concerning short GRBs, from the absence of evidence of simultaneous SN explosions  and from the association with galaxies with a wide range of star formation properties (inclusive of low SFR), it can be concluded that, very likely, they are the result of compact binary (e.g., NS-NS) merging, as it was initially supposed. The recent association of a gravitational wave signal (GW~170817) with a short GRB (170817), has confirmed this scenario \cite{Abbott17}.

Ruffini et al.\cite{Ruffini16} have proposed a binary nature also for the progenitors of long XRFs/GRBs which exhibit two distinct episodes in their light curves. According to their model, in these cases a CO$_{core}$  undergoes a SN explosion which triggers an hypercritical accretion onto a NS companion in a tight or more separated binary system. Depending on the tightness level, the formation or not of a BH is driven, and a GRB or an XRF event, respectively, is produced. An outstanding candidate they consider for their model in the case of formation of a black hole is the very luminous GRB~090618 ($E_{iso} = 3 \times 10^{53}$~erg) for which there is an evidence of a SN bump 10 days after the event\cite{Izzo12}, while an example of the second class of candidate events is the underluminous GRB~060218 ($E_{iso} = 5 \times 10^{49}$~erg), also with an associated SN and the evidence of a SN shock breakout (see \cite{Waxman07} and Table~\ref{t:SNe-GRBs}).

\section{Some still open issues and opportunities offered by GRBs}

Several questions are still open on GRB physics and properties, like the central engine that powers the GRB events (black hole plus torus? a magnetar? a quark star?), the ejecta composition (matter dominated? magnetically-dominated jet?), the radiation mechanism (synchrotron radiation? synchrotron self Compton? thermal upscattering of a thermal photon source?), the still not measured hard X--ray afterglow spectrum (is it the tail of that measured at low energies?), cosmological issues (are GRBs good tracers of the star formation history of the Universe? Can high-z GRBs probe the reionization history of the Universe? Are GRBs sources of ultra high-energy gamma rays? If yes, which is the emission physics?). For a review of the open issues see, e.g., \cite{Zhang11a}.

GRBs also offer the opportunity to settle questions of fundamental physics, like the test of the Lorentz invariance violation that is expected in some theories of quantum gravity. As above mentioned, GRBs can also be used as beacons to derive the cosmological parameters, and they are crucial for the multi-messenger astronomy \cite{Abbott17}.  

It is expected that many other exciting discoveries will be done in the next future with space and ground facilities, some of which  already operational like \swift, \fermi, \agile, and \integral, the large optical and radio telescopes like VLT, Keck, EVLA, and the  LIGO and VIRGO gravitational interferometers. 

\begin{figure}
	\centering\includegraphics[width=0.6\textwidth]{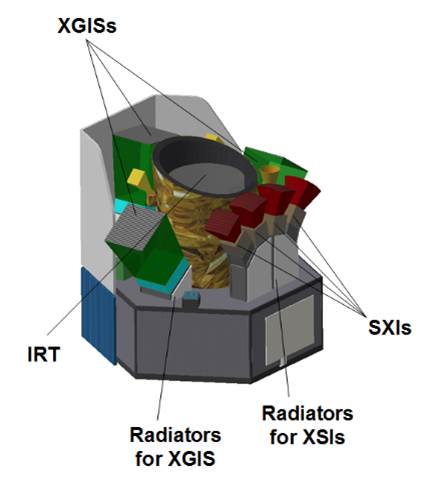}
		\caption[]{The THESEUS mission, selected by ESA for a Phase A study. If adopted, its launch is foreseen in 2032. Reprinted from \cite{Amati18}.}
\label{f:theseus}
\end{figure}

In addition, other missions are being studied, or are under development, or are just operational.  
They include space X--/gamma--ray missions, like the Chinese-French mission {\em SVOM} (4~keV--5~MeV) expected to be launched in 2021, the Chinese-UK mission {\em Einstein Probe} (0.5--4~keV), aimed for a launch by the end of 2022,  the {\em THESEUS} mission just approved by ESA for a phase A study (see Fig.~\ref{f:theseus}), with a launch in 2032 if approved; extremely large optical facilities like {\em EELT}, {\em TMT} and {\em GMT};    
new radio facilities like {\em LOFAR} and {\em SKA};  gravitational wave missions like the large ESA mission {\em eLISA} foreseen to be launched in the early 2030s; very high-energy gamma--ray facilities like {\em MAGIC}, {\em HESS}, {\em VERITAS}, and {\em CTA}; large neutrino facilities like {\em ANTARES}, {\em ICECUBE}, and {\em KM3NET}.

%
% Figure 14
%

I would like to conclude with an X--/gamma--ray mission concept, {\em ASTENA} (Advanced Surveyor of Transient Events and Nuclear Astrophysics) under study by an international collaboration led by the University of Ferrara in the framework of the European programme AHEAD. The mission includes an array of 18 Wide Field Monitors-Imager Spectrometers (WFM-IS) with a total useful area of about 2 m$^2$ and an energy band from 2~keV to 
20~MeV, and a focusing Narrow Field Telescope (NFT) with a collection area of about 7~m$^2$ and an energy band from 50 keV to 700 keV. Its sensitivity is expected to improve that of the best gamma--ray instruments, inclusive of NuSTAR, by 2 orders of magnitude. The NFT will be the ideal instrument to study, among others, the high-energy GRB afterglow spectra still unknown.

In conclusion, the discovery of the GRB sites and afterglow with \sax\ will continue to push up the frontiers
of our knowledge of the Universe for many years.

\bibliographystyle{spmpsci}      % mathematics and physical sciences
\bibliography{grb_ref}   % name your BibTeX data base

\end{document}